\documentclass[conference,letterpaper]{IEEEtran}
\usepackage[utf8]{inputenc}
\usepackage{fancyhdr}
\usepackage{amsmath}%
\usepackage{graphicx}

\usepackage{cancel}

\usepackage{amsmath} 
\usepackage{amssymb} 
\usepackage{pifont} 
 \usepackage{cite}
\usepackage{url}
\usepackage{comment}
\usepackage{color, colortbl}
\usepackage{enumerate}
\usepackage{arcs}

\usepackage[framemethod=TikZ]{mdframed}

\usepackage{fancyhdr}

\usepackage{pdfpages}

\newtheorem{defini}{Definition}
\newtheorem{thm}{Theorem}

\newtheorem{exemple}{Example}

\newtheorem{lem}{Lemma}
\definecolor{Gray}{gray}{0.9}
\begin{document}


\title{{A Theorem for Secrecy in Tagged Protocols Using the Theory of Witness-Functions}}




\author{\IEEEauthorblockN{Jaouhar Fattahi}

\IEEEauthorblockA{Valcartier Research Centre. Defence Research and Development Canada.\\ 2459 de la Bravoure Road, Québec, Canada. G3J 1X5.}
\IEEEauthorblockA{Laboratory of Computer Security. Department of Computer Science and Software Engineering.\\ Pavillon Adrien-Pouliot, local 3770. Laval University. Québec, Canada, G1K 7P4. \\ E-mail: jaouhar.fattahi@drdc-rddc.gc.ca $\vert$ jaouhar.fattahi.1@ulaval.ca}


}


%


\maketitle
\thispagestyle{plain}

\fancypagestyle{plain}{
\fancyhf{}	
\fancyfoot[L]{}
\fancyfoot[C]{}
\fancyfoot[R]{}
\renewcommand{\headrulewidth}{0pt}
\renewcommand{\footrulewidth}{0pt}
}

\pagestyle{fancy}{
\fancyhf{}
\fancyfoot[R]{}}
\renewcommand{\headrulewidth}{0pt}
\renewcommand{\footrulewidth}{0pt}

\begin{abstract}

In this paper, we enunciate the theorem of secrecy in tagged protocols using the theory of witness-functions and we run a formal analysis on a new tagged version of the Needham-Schroeder public-key protocol using this theorem. We discuss the significance of tagging in securing cryptographic protocols as well. \\
                                                                                                                                                                                                                                                                                                                                                                                                                                                                                                                                                                                                                                                                                                                                                                                                                                                                                                                                                                                                                                                                                                                                                                                                                                                                                                                                                                                                                                                                                                                                                                                                                                                                                                                                                                                                                                                                                                                                                                                                                                                                                                                                                                                                                                                                                                                                                                                                                                                                                                                                                                                                                                                                                                                                                                                                                                                                                                                                                                                                                                                                                                                                                                                                                                                                                                                                                                                                                                                                                                                                                                                                                                                                                                                                                                                                                                                                                                                                                                                                                                                                                                                                                                                                                                                                                                                                                                                                                                                                                                                                                                                                                                                                                                                                                                                                                                                                                                                                                                                                                                                                                                                                                                                                                                                                                                                                                                                                                                                                                                                                                  
\end{abstract}


%
\IEEEpeerreviewmaketitle

\section*{Notice \footnote{\label{myfootnote}This paper has been accepted at the 31st Annual IEEE Canadian Conference on Electrical and Computer Engineering (CCECE 2018). Québec City, Canada. May 13–16, 2018.}}

\textit{© 2018 IEEE. Personal use of this material is permitted. Permission from IEEE must be obtained for all other uses, in any current or future media, including reprinting/republishing this material for advertising or promotional purposes, creating new collective works, for resale or redistribution to servers or lists, or reuse of any copyrighted component of this work in other works.}

\section{Introduction}

Recently, a new category of analytic functions, called witness-functions,                                                                                                                                                                                                                                                                                                                                                                                                                                                                                                                                                                                                                                                                                                                                                                                                                                                                                                                                                                                                                                                                                                                                                                                                                                                                                                                                                                                                                                                                                                                                                                                                                                                                                                                                                                                                                                                                                                                                                                                                                                                                                                                                                                                                                                                                                                                                                                                                                                                                                                                                                                                                                                                                                                                                                                                                                                                                                                                                                                                                                                                                                                                                                                                                                                                                                                                                                                                                                                                                                                                                                                                                                                                                                                                                                                                                                                                                                                                                                                                                                                                                                                                                                                               has been put forward to analyze cryptographic protocols for secrecy\cite{ChapterJFSpringer,TheseJF,8123025,DBLP:conf/apn/FattahiMH14}. These functions assign to every single atomic message involved in the protocol a reasonable level of security. An analysis with a witness-function is the process that tries to make sure that this level of security never goes down between any two consecutive steps, a receiving step and a sending one, from the very first appearance of the atomic message in the protocol until its final destination. This is obviously sufficient to guarantee that any secret will never fall into the hands of an unauthorized agent including an evil intruder. In that case, the protocol is said to be increasing. Certainly, the witness-functions are able to analyze any protocol. However, we notice that they present interesting features when they are used on tagged protocols. In fact, the theorem of analysis acquires a reduced and elegant form and the analysis becomes much quicker. This is because there is a subtle relationship between tagging, on the one hand, and a witness-function definition, on the other hand. In this paper, we discuss these aspects and we analyze a tagged protocol with a witness-function.                                                                                                                                                                                                                                                                                                                                                                                                                                                                                                                                                                                                                                                                                                                                                                                                                                                                                                                                                                                                                                                                                                                                                                                                                                                                                                                                                                                                                                                                                                                                                                                                                                                                                                                                                                                                                                                                                                                                                                                                                                                                                                                                                                                                                                                                                                                                                                                                                                                                                                                                                                                                                                                                                                                                                                                                                                                                                                                                                                            The paper is organized as follows. In section \ref{sec1}, we recall the theory of witness-functions. In section \ref{sec2}, we give an overview on tagged protocol. In section \ref{sec3}, we enunciate the theorem of secrecy in tagged protocols using witness-functions. In section \ref{sec4}, we propose a tagged version of the Needham-Schroeder  public-key protocol and we analyze it with that theorem. In section \ref{sec5}, we discuss some interesting related works dealing with tagged protocols and we compare them to our approach. In section \ref{sec6}, we conclude.

\section{The theory of witness-functions}\label{sec1}

The theory of witness-functions  has been proposed by Fattahi et al. \cite{ChapterJFSpringer,TheseJF,8123025,DBLP:conf/apn/FattahiMH14} to statically verify cryptographic protocols for secrecy. A witness-function is an analytic function that attributes a safe level of security to                                                                                                                                                                                                                                                                                                                                                                                                                                                                                                                                                                                                                                                                                                                                                                                                                                                                                                                                                                                                                                                                                                                                                                                                                                                                                                                                                                                                                                                                                                                                                                                                                                                                                                                                                                                                                                                                                                                                                                                                                                                                                                                                                                                                                                                                                                                                                                                                                                                                                                                                                                                                                                                                                                                                                                                                                                                                                                                                                                                                                                                                                                                                                                                                                                                                                                                                                                                                                                                                                                                                                                                                                                                                                                                                                                                                                                                                                                                                                                                                                                   every atomic message in the protocol and the analysis using a witness-function closely follows                                                                                                                                                                                                                                                                                                                                                                                                                                                                                                                                                                                                                                                                                                                                                                                                                                                                                                                                                                                                                                                                                                                                                                                                                                                                                                                                                                                                                                                                                                                                                                                                                                                                                                                                                                                                                                                                                                                                                                                                                                                                                                                                                                                                                                                                                                                                                                                                                                                                                                                                                                                                                                                                                                                                                                                                                                                                                                                                                                                                                                                                                                                                                                                                                                                                                                                                                                                                                                                                                                                                                                                                                                                                                                                                                                                                                                                                                                                                                                                                                                                                                                                                                                                                        the growth of this value during the lifecycle of this atom. In this                                                                                                                                                                                                                                                                                                                                                                                                                                                                                                                                                                                                                                                                                                                                                                                                                                                                                                                                                                                                                                                                                                                                                                                                                                                                                                                                                                                                                                                                                                                                                                                                                                                                                                                                                                                                                                                                                                                                                                                                                                                                                                                                                                                                                                                                                                                                                                                                                                                                                                                                                                                                                                                                                                                                                                                                                                                                                                                                                                                                                                                                                                                                                                                                                                                                                                                                                                                                                                                                                                                                                                                                                                                                                                                                                                                                                                                                                                                                                                                                                                                                                                                                                                                                                        section, we recall the                                                                                                                                                                                                                                                                                                                                                                                                                                                                                                                                                                                                                                                                                                                                                                                                                                                                                                                                                                                                                                                                                                                                                                                                                                                                                                                                                                                                                                                                                                                                                                                                                                                                                                                                                                                                                                                                                                                                                                                                                                                                                                                                                                                                                                                                                                                                                                                                                                                                                                                                                                                                                                                                                                                                                                                                                                                                                                                                                                                                                                                                                                                                                                                                                                                                                                                                                                                                                                                                                                                                                                                                                                                                                                                                                                                                                                                                                                                                                                                                                                                                                                                                                                                                                                          fundaments of this theory.                                                                                                                                                                                                                                                                                                                                                                                                                                                                                                                                                                                                                                                                                                                                                                                                                                                                                                                                                                           Please	notice that we will give the meaning of every notation we use in a natural language as soon as it shows up first. 

\subsection{Context of verification}

A protocol analysis using the witness-functions runs in a role-based specification\cite{DBLP:conf/acsac/DebbabiLM98, Mejri} under the hypotheses of Dolev-Yao\cite{1056650}. In this paper, we assume that a protocol is always analyzed under the perfect encryption assumption which means that we do not deal with flaws caused by the cryptographic system in use or the implementation of cryptographic primitives. Equally, we suppose that there is no special equational theory and all secrets, keys and other names are atomic.

\subsection{Reliable function}
\begin{defini}{(Well-formed Function)}\label{bienforme}
{
Let ${F}$ be a function. ${F}$ is well-formed iff:
$\forall M,M_1,M_2 \subseteq {\mathcal{M}}, \forall \alpha \in {\mathcal{A}}({\mathcal{M}}) \mbox{:}$

\begin{tabular}{lll}
   ${F}(\alpha,\{\alpha\})$ & $=$ & $\bot$ \\
   ${F}(\alpha, {M}_1 \cup {M}_2)$ & $=$ & ${F}(\alpha, {M}_1)\sqcap{F}(\alpha,{M}_2)$ \\
   ${F}(\alpha,{M})$ & $=$ & $\top, \mbox{ if } \alpha \notin {\mathcal{A}}({M})$ \\
\end{tabular}
}$ $\\
\end{defini}

A well-formed function ${F}$ should assign the infimum level of security (i.e. $\bot$) to an atomic message $\alpha$ that shows up in clear (not encrypted) in a set $M$ of messages. This is obviously to express that  anybody who knows $M$ inevitably knows $\alpha$. It  assigns to an atomic message in the union of two sets of messages the minimum of the two levels (i.e. $\sqcap$) assigned in each set alone. It assigns the supremum (i.e. $\top$) to an atomic message $\alpha$ that does not even appear in $M$. This is to express the fact that nobody is able to know $\alpha$ when he knows $M$. We note by ${\mathcal{A}}({M})$ the atomic messages of $M$.

\begin{defini}{(Full-invariant-by-intruder Function)}\label{spi}
{
Let ${F}$ be a function. 
${F}$ is full-invariant-by-intruder iff: $\forall {M} \subseteq {\mathcal{M}}, m\in {\mathcal{M}},  \alpha \in {\mathcal{A}}(m)$:\\
$
 {M} \models m \Rightarrow ({F}(\alpha,m) \sqsupseteq{F}(\alpha,{M})) \vee (\ulcorner K(I) \urcorner \sqsupseteq \ulcorner \alpha \urcorner).
$
}
\end{defini}

A full-invariant-by-intruder function ${F}$ should resist against any malicious tentative to lower the level of security by an intruder once $F$ assigns to an atomic message $\alpha$ a level of security in a set of messages $M$. That is to say that the intruder can never infer (i.e.   
 $\models$) from this set $M$ any other message $m$ in which this level may be lower than the one given in $M$ (i.e. ${F}(\alpha,m) \not \sqsupseteq{F}(\alpha,{M})$), exception made when the intruder is explicitly authorized to know $\alpha$ (i.e. $\ulcorner K(I) \urcorner \sqsupseteq \ulcorner \alpha \urcorner$). We say that a function ${F}$ is reliable when it is well-formed and full-invariant-by-intruder.

\begin{defini}{(${F}$-Increasing Protocol)}\label{ProAbsCroi}
{
Let ${F}$ be a function and $p$ be a protocol. $p$ is ${F}$-increasing iff: $\forall R.r, \forall \sigma,  \forall \alpha \in {\mathcal{A}}(r^+), \mbox{ we have: }{F}(\alpha, r^+\sigma)\sqsupseteq \ulcorner \alpha \urcorner \sqcap{F}(\alpha, R^-\sigma)$
}
\end{defini}
$ $\\
An ${F}$-increasing protocol is a protocol that constantly pumps traces (substituted generalized roles in a role-based specification) with atomic messages $\alpha$ that always have a security level, calculated by ${F}$, higher (i.e. $\sqsupseteq$) upon a sending step (i.e. in the generalized role $ r^+\sigma$, the sign $+$ denotes a sending operation and $\sigma$ a substitution corresponding to a possible execution of the protocol) than the one calculated by the same function in the messages received in the latest receiving step (i.e. in the generalized role $R^-\sigma$, the sign $-$ denotes a receiving operation), or higher than the level of security of $\alpha$ obtained directly from within the context of verification (i.e. $\ulcorner \alpha \urcorner$), if it is available. \\

\begin{thm}{(Secrecy in Increasing Protocols)}\label{mainTh}
{
Let ${F}$ be a reliable function and $p$ be an ${F}$-increasing protocol.
\begin{center}
$p$ is correct for secrecy.
\end{center}
}
\end{thm}

Theorem \ref{mainTh} brings up a very important result. It establishes that a protocol is correct for secrecy if it could be demonstrated increasing using a reliable function $F$. The sketch of the proof is quite straightforward. That is, if the attacker manages to discover an initially protected secret $\alpha$ (get it in clear) then its security level calculated  by $F$ should be the infimum seeing as $F$ is well-formed. This scenario cannot be rooted in the rules of the protocol seeing as this latter is $F\mbox{-increasing}$ and its rules constantly raise the level of security of $\alpha$. This scenario could not happen either if the intruder uses his capabilities seeing as $F$ is full-invariant-by-intruder and then the intruder could not forge any message in which the security level of $\alpha$ may decline. Hence, this scenario could simply never happen and the secret could never be disclosed. The complete formal proof could be found in~\cite{Relaxed}.

\subsection{Construction of Reliable Function}\label{sectionFonctionsetSelections}

Here we give one constructive way to build a reliable function. Let's consider the function $F$ defined as follows:\\

\begin{defini}{(Reliable Function)}\label{reliable}
\scalebox{0.9}{
\begin{tabular}{llcl}
 1.& ${F}(\alpha,\{\alpha\})$ & $=$ & $\bot$ \\
  2. & ${F}(\alpha, {M}_1 \cup {M}_2)$ & $=$ & ${F}(\alpha, {M}_1)\sqcap{F}(\alpha,{M}_2)$ \\
   3. & ${F}(\alpha,{M})$ & $=$ & $\top, \mbox{ if } \alpha \notin {\mathcal{A}}({M})$ \\
   4. & ${F}(\alpha,m_1.m_2)$ & $=$ & $F(\alpha,\{m_1,m_2\})$ \\
   5. & ${F}(\alpha,\{m\}_{k})$ & $=$ & $F(\alpha,\{m\}),$ \mbox{ if } $\ulcorner k^{-1} \urcorner \not \sqsupseteq \ulcorner \alpha \urcorner$\\
   6. & ${F}(\alpha,\{m\}_{k})$ & $=$ & $\ulcorner k^{-1} \urcorner  \sqcap \mbox{ID}(m),$ \mbox{ if } $\ulcorner k^{-1} \urcorner \sqsupseteq \ulcorner \alpha \urcorner$\\
\end{tabular}
}
\end{defini}
$ $\\
The first three steps 1., 2. and 3. directly grant the function $F$ the property of being well-formed. The step 4. deconcatenates  a message $m_1.m_2$ into two messages $m_1$ and $m_2$ and $F$ returns the same level of security as in the set $\{m_1,m_2\}$. That is because an intruder, although he can deconcatenate any message $m_1.m_2$, he cannot infer about $\alpha$ in $m_1.m_2$ more than he could infer about it in each of $m_1$ or $m_2$ separately. The step 5. ignores encryption with an outer weak key (i.e. $\ulcorner k^{-1} \urcorner \not \sqsupseteq \ulcorner \alpha \urcorner$) and looks for a deeper strong key. That is because if $\alpha$ is encrypted with a weak key, it can fall into the hands of an unauthorized agent. The step 6. makes sure that $\alpha$ is encrypted with a strong key $k$ (i.e. $\ulcorner k^{-1} \urcorner \sqsupseteq \ulcorner \alpha \urcorner$ meaning the reverse key $k^{-1}$ must be known only by a part of agents who are authorized to know $\alpha$ in the context) and $F$ returns the set of agent identities who know the reverse key (i.e. $\ulcorner k^{-1} \urcorner$) as well as the identity of all the neighbors of $\alpha$ in $m$ (i.e. $\mbox{ID}(m)$). The step 6. transforms $F$ into a full-invariant-by-intruder function. In fact, an unauthorized intruder who attempts to mislead $F$ should obtain the key $ k^{-1}$ beforehand. Hence,  his knowledge must include $k^{-1}$ (i.e. $\ulcorner K(I) \urcorner \sqsupseteq \ulcorner k^{-1} \urcorner$). Since the key $ k^{-1}$ is such that $\ulcorner k^{-1} \urcorner \sqsupseteq \ulcorner \alpha \urcorner$ then the knowledge of the intruder must satisfy $\ulcorner K(I) \urcorner \sqsupseteq \ulcorner \alpha \urcorner$ as well owing to the transitivity of the comparator "$\sqsupseteq$". This is contradictory to the fact that the intruder is unauthorized to know $\alpha$.

\begin{exemple}
Let us have the following context of verification: $\ulcorner \alpha \urcorner=\{A, B, S\}$; $m=\{\{C.\{\alpha.D\}_{k_{as}}\}_{k_{ab}}\}_{k_{ac}}$; ${k_{ac}^{-1}}={k_{ac}},{k_{ab}^{-1}}={k_{ab}}, {k_{as}^{-1}}={k_{as}}$; $\ulcorner{k_{ac}}\urcorner=\{A, C\},\ulcorner{k_{as}}\urcorner=\{A, S\}, \ulcorner{k_{ab}}\urcorner=\{A, B\}$. We have: \\
$F(\alpha,m)=F(\alpha,\{C.\{\alpha.D\}_{k_{as}}\}_{k_{ab}})=\{C, D\}{ \cup}\ulcorner{k_{ab}^{-1}}\urcorner=\{C,D\} \cup \{A,B\}=\{A, B, C, D\}$.\\ Please notice that the outermost encryption by ${k_{ac}}$ has been ignored by $F$ because it is a weak key since the agent $C$ is not authorized to know $\alpha$ in the context (i.e. $\ulcorner \alpha \urcorner=\{A, B, S\}$). This case falls into the step 5.
\end{exemple} 

Other reliable functions could be found in \cite{DBLP:conf/apn/FattahiMH14,TheseJF}. In the rest of this paper, we will only use the function defined in this subsection and we refer to it by $F$. 

\subsection{Witness-functions to reduce the impact of variables}\label{sectionWF}

The function $F$ as defined above may be suitable to assign security level for atomic messages but in ground terms only. Nevertheless, when we analyze a protocol, messages are not necessarily ground and may contain variables. To cope with this situation, the idea is to use the derivative function $F'$ of $F$ that operates like $F$ but after eliminating variables from the neighborhood of $\alpha$ (e.g. $F'(\alpha,\{\alpha.X.B\}_{k_{cd}})=F(\alpha,\{\alpha.{B}\}_{k_{{cd}}})=\{B, C, D\}$). Although the derivative function remains well-formed and full-invariant-by-intruder, it may lose its quality as a function and may return multiple and contradictory values for the same trace generated by a substitution in the generalized roles. For example, if the trace is $\{\alpha.A.{B}\}_{k_{{cd}}}$ that could be produced by substitution in two generalized roles $\{\alpha.X.{B}\}_{k_{{cd}}}$ and $\{\alpha.Y\}_{k_{{cd}}}$, the function $F'$ assigns to $\alpha$ the level of security $\{B, C, D\}$ when the trace originates from the first generalized role, and the level of security $\{C, D\}$ if the trace originates from the second one. To overcome this incoherence, we define the witness-functions.

\begin{defini}{[Witness-Function]}\label{WF}
\[{{{\mathcal{W}}}}_{p,F}(\alpha,m\sigma)=\underset{{\{(m',\sigma') \in \tilde{\mathcal{M}}_p^{\mathcal{G}}\otimes\Gamma|m'\sigma' = m \sigma \}}}{\sqcap} \!\!\!\!\!\!\!\!\!\!\!\!\!\!\!\!\!\!\!\!\!\!\!\!\!\!\!F'(\alpha, m'\sigma')\]
\end{defini}

A witness-function ${{{\mathcal{W}}}}_{p,F}$ calculates the level of security of an atomic message $\alpha$ in a trace $m\sigma$ by using $F'$ applied to all the possible origins $m'$ in the messages $\tilde{\mathcal{M}}_p$ generated by the generalized roles and returns the minimum, which is obviously a single value. Nevertheless, a witness-function could not be used as is to analyze a protocol since the analysis runs statically on the generalized roles not on the traces (i.e. $m\sigma$) which are dynamic entities. For that, we bound a witness-function by two static bounds as follows.

\begin{lem}{[binding a witness-function]}\label{prePAT}
$$F'(\alpha, m)\sqsupseteq {{{\mathcal{W}}}}_{p,F}(\alpha,m\sigma)\sqsupseteq \!\!\!\! \!\!\!\! \!\!\! \underset{{\{(m',\sigma') \in \tilde{\mathcal{M}}_p^{\mathcal{G}}\otimes\Gamma|m'\sigma' = m \sigma' \}}}{\sqcap} \!\!\!\!\!\!\!\!\!\!\!\!\!\!\!\!\!\!\!\!\!\!\!\!\!\!\!\!F'(\alpha, m'\sigma')$$
\end{lem}

The upper-bound $F'(\alpha, m)$ returns a minimal set of identities from $m$ after removing all variables in $m$. The lower-bound $\underset{{\{(m',\sigma') \in \tilde{\mathcal{M}}_p^{\mathcal{G}}\otimes\Gamma|m'\sigma' = m \sigma' \}}}{\sqcap} \!\!\!\!\!\!\!\!\!\!\!\!\!\!\!\!\!\!\!\!\!\!\!\!\!\!\!\!F'(\alpha, m'\sigma')$ returns all the identities gathered from all the messages that could be unified with $m$. The witness-function returns certain identities in between which are known only when the protocol is executed from the actual origins of the trace only. The inequality is quite intuitive since $m$ is a guaranteed origin of the trace $m\sigma$ and the actual origins of the trace $m\sigma$ is a subset of the messages that are unifiable with $m$. The two bounds are obviously statically computable. 

\begin{thm}{[Decision Procedure for Secrecy with a Witness-Function]}\label{PAT}
Let $p$ be a protocol. Let ${\mathcal{W}}_{p,F}$ be a witness-function.
$p$ is correct for secrecy if:
$\forall R.r \in R_G(p), \forall \alpha \in {\mathcal{A}}{(r^+ )}$ we have:
$$\underset{{\{(m',\sigma') \in \tilde{\mathcal{M}}_p^{\mathcal{G}}\otimes\Gamma|m'\sigma' = r^+ \sigma' \}}}{\sqcap} \!\!\!\!\!\!\!\!\!\!\!\!\!\!\!\!\!\!\!\!\!\!\!\!\!\!\!\!\!F'(\alpha,  m'\sigma') \sqsupseteq \ulcorner \alpha \urcorner \sqcap F'(\alpha, R^-)$$
\end{thm}

Theorem \ref{PAT} establishes a decision procedure for secrecy using the bounds a witness-function. When a message is sent (i.e. $r^+$), it is analyzed largely with the lower-bound of a witness-function. When a message is received (i.e $R^-$), it is analyzed strictly with the upper-bound of a witness-function. Any dishonest identity ambushed by the lower-bound that is not returned by the upper-bound will be interpreted as an intrusion. The protocol is then decided not increasing and the analysis halts with a failure flag. Theorem \ref{PAT} is a direct result of Theorem \ref{mainTh} and Lemma \ref{prePAT}. Please notice that Theorem \ref{PAT} does not imply the witness-function itself (i.e. ${{{\mathcal{W}}}}_{p,F}$). It involves its bounds only.

\section{Tagged Protocols}\label{sec2}

A tag is any subtlety or any syntactic annotation put inside a message to differentiate it from another message. A tagged protocol is a protocol such that every message received by any agent has a unique and regular origin. That implies that every single message (an encryption pattern) containing a variable (i.e. something that the receiver does not know) is distinguishable from any other message (any other encryption pattern) and does not unify with any message other than the regular message that the receiver is expecting to get through the network from the right agent. Tagging a protocol could be reached by inserting an identity beside some atom in the message. For example, if an agent $A$ receives the message $\{\alpha.B.X\}_{k_{ab}}$ where the variable $X$ is supposed to be a nonce $N_b$ sent by a regular agent $B$ and the protocol generates also the message $\{\alpha.B.C\}_{k_{ab}}$, we can change the message $\{\alpha.B.X\}_{k_{ab}}$ in the definition of the protocol by $\{A.\alpha.B.X\}_{k_{ab}}$ in the new tagged version of the protocol and hence the message $\{\alpha.B.C\}_{k_{ab}}$ will not unify with it. All the same, a signature could be an efficient tag, too. For example, we can change the message $\{\alpha.B.X\}_{k_{ab}}$ in the definition of the protocol by $\{\{\alpha\}_{k_{b}^{-1}}.B.X\}_{k_{ab}}$ in the new tagged version of the protocol to prevent $\{\alpha.B.C\}_{k_{ab}}$ from unifying with it. Tagging could be also reached by inserting an ordinal number into an encrypted message or inserting a string describing the type of certain components inside. In general, tagging prevents man-in-the-middle attacks from happening by offering the receiver the way to distinguish a regular message from an irregular one. 

\section{Theorem for Secrecy in Tagged Protocols} \label{sec3}

As a matter of fact, when a protocol is tagged (all its messages are distinguishable one from another), it becomes nonsense talking about message that overlap (unifiable). This has a direct impact on the reduction of Theorem \ref{PAT}. In fact, the expression $\underset{{\{(m',\sigma') \in \tilde{\mathcal{M}}_p^{\mathcal{G}}\otimes\Gamma|m'\sigma' = r^+ \sigma' \}}}{\sqcap} \!\!\!\!\!\!\!\!\!\!\!\!\!\!\!\!\!\!\!\!\!\!\!\!\!\!\!\!\!F'(\alpha,  m'\sigma')$ in Theorem \ref{PAT} will be reduced to $F'(\alpha,r^+)$. That is, the lower-bound $\underset{{\{(m',\sigma') \in \tilde{\mathcal{M}}_p^{\mathcal{G}}\otimes\Gamma|m'\sigma' = r^+ \sigma' \}}}{\sqcap} \!\!\!\!\!\!\!\!\!\!\!\!\!\!\!\!\!\!\!\!\!\!\!\!\!\!\!\!\!F'(\alpha,  m'\sigma')$ means $F'$ applied to all the patterns in the generalized roles that are unifiable with the message $r^+$ and its goal is to ambush dishonest identities that could be inserted in the neighborhood of an analyzed atomic message $\alpha$. Nevertheless, this could never happen when the protocol is tagged. A tagged protocol creates in fact a series of \textit{from regular to regular}  data flow in which the intruder is hopeless to launch any man-in-the-middle attack. In that case, if the protocol happens to be incorrect, that will definitely be because it is not increasing by construction because of a bad reasoning on the knowledge of every agent and without any intervention from the intruder. This brings us to the following theorem.\\

\begin{thm}{[Theorem of Secrecy for Tagged Protocols]}\label{PATTag}
Let $p$ be a tagged protocol. Let ${\mathcal{W}}_{p,F}$ be a witness- function.
$p$ is correct for secrecy if:
$\forall R.r \in R_G(p), \forall \alpha \in {\mathcal{A}}{(r^+ )}$ we have:
$$F'(\alpha,r^+) \sqsupseteq \ulcorner \alpha \urcorner \sqcap F'(\alpha, R^-)$$
\end{thm}

Theorem \ref{PATTag} enables tagged protocols to use simply the derivative function $F'$ on both the received generalized role and the sent one to determine whether or not the tagged protocol is increasing, with no need to perform any further unifications. It is worth mentioning that verifying whether or not a protocol is tagged is an easy task that is carried out only once before analyzing the protocol. It is equally worth noticing that Theorem \ref{PATTag} sets just sufficient conditions for the tagged protocol correctness regarding secrecy, which conditions are not inevitably necessary since the problem of secrecy remains undecidable in general. In the rest of the paper, we will refer to Theorem \ref{PATTag} by the acronym TSTP.


\section{Formal analysis of a tagged version of the Needham-Schroeder  public-key protocol}\label{sec4}
In this section, we propose our new tagged version of the Needham-Schroeder public-key protocol (different from the NSL protocol) and we analyze it with Theorem \ref{PATTag} (TSTP) for secrecy. This version is given in Table  \ref{NSLVar}.
\begin{table}
\begin{center}
\begin{tabular}{|llllll|}
\hline
 $p$= &$\langle1,$ & $A$ & $\longrightarrow$ & $B:$ & $\{N_a.A.B\}_{k_b}\rangle$ \\
  &$\langle2,$ & $B$ & $\longrightarrow$ & $A:$ & $\{A.B.N_a\}_{k_a}.\{B.A.N_b\}_{k_a}\rangle$ \\
  &$\langle3,$ & $A$ & $\longrightarrow$ & $B:$ & $\{N_b.B.A.N_a\}_{k_b}\rangle.$\\
\hline
\end{tabular}
\end{center}
\caption{a tagged version of the  Needham-Schroeder protocol}
\label{NSLVar}
\end{table}

\subsection{Context setting}

The generalized roles of $p$ are defined by $\cal{R}_{\cal{G}}(\textit{p})=\{A_{\cal{G}},B_{\cal{G}}\}$ where:

\small{
\begin{center}
\scalebox{0.9}{
\begin{tabular}{lllllll}
  $\cal{A}_{\cal{G}}=$ & $i.1$ & $A$& $\longrightarrow I(B)$&:& $\{N_a^i.A.B\}_{k_b}$ \\
  $$ & $i.2$ & $I(B)$&$ \longrightarrow A$ & : & $\{A.B.N_a^i\}_{k_a}.\{B.A.X\}_{k_a}$ \\
  $$ & $i.3$ & $A$& $\longrightarrow I(B)$ & : & $\{X.B.A.N_a^i\}_{k_b}$ \\
  $\cal{B}_{\cal{G}}=$ & $j.1$ & $I(A)$& $\longrightarrow B$ & : & $\{Y.A.B\}_{k_b}  $\\
  $$ & $j.2$ & $B$&$ \longrightarrow I(A)$ & : & $\{A.B.Y\}_{k_a}. \{B.A.N_b^j\}_{k_a}$ \\
  $$ & $j.3$ & $I(A)$&$ \longrightarrow B$ & : & $\{N_b^j.B.A.Y\}_{k_b}$ 
\end{tabular}
}
\end{center}
}
$ $\\
Initial knowledge :\\ 
$\ulcorner A\urcorner= \bot$; $\ulcorner B\urcorner= \bot$; (i.e. two public identities)\\
$\ulcorner N_a\urcorner= \{A,B\}$ (i.e. secret shared between $A$ and $B$);\\ $\ulcorner N_b\urcorner= \{A,B\}$ (i.e. secret  shared between $A$ and $B$); \\
$\ulcorner k_a^{-1}\urcorner=\{A\}$; (i.e. private key of $A$)\\
$\ulcorner k_b^{-1}\urcorner=\{B\}$; (i.e. private key of $B$)\\
$\ulcorner k_a \urcorner=\bot$; (i.e. public key of $A$)\\
$\ulcorner k_b \urcorner=\bot$; (i.e. public key of $B$)\\
$({\cal{L}},\sqsupseteq, \sqcup, \sqcap, \bot,\top)=(2^{\cal{I}},\subseteq,\cap,\cup,\cal{I}, \emptyset)$; (i.e. security lattice)\\
${\cal{I}}=\{I, A, B\}$; (i.e. intruder and regular agents present on the net)\\
${\cal{X}}_p=\{X, Y\}$ is the set of variables. $F$ is the function given by Definition \ref{reliable} and $F'$ is its derivative form. 

\subsection{Tagging verification}

Before we dive into the analysis, let us make sure that this protocol is a tagged one. At the first sight, an attentive eye should remark that the protocol is tagged by the position of the identities in its messages. In fact, the encrypted message $\{N_a.A.B\}_{k_b}$ is the only one that contains the identity of the receiver (i.e. $B$) at the last position. The encrypted message $\{A.B.N_a\}_{k_a}$ is the only one that contains the identity of the receiver (i.e. $A$) at the first position and the identity of the sender (i.e. $B$) at the second position. The encrypted message $\{B.A.N_b\}_{k_a}$ is the only one that contains the identity of the receiver (i.e. $A$) at the second position and the identity of the sender (i.e. $B$) at the first position. Finally, the encrypted message $\{N_b.B.A.N_a\}_{k_b}$ is the only one that contains the identity of the receiver (i.e. $B$) followed by the identity of the sender (i.e. $A$) that must show up in the middle of the message only. This makes all the encryptions distinguishable one from another from a receiver point of view. More rigorously, according to the generalized roles $A_{\cal{G}}$, the agent $A$ is a receiver in the step $i.2$. The first message he receives is $\{A.B.N_a^i\}_{k_a}$ which is the regular message expected by $B$. The other message is $\{B.A.X\}_{k_a}$ which unifies only with $\{B.A.N_b^i\}_{k_a}$ which is the regular message that $A$ is expecting.\\

According to the generalized roles $B_{\cal{G}}$, the agent $B$ is a receiver in two steps.
\begin{enumerate}
\item In the step $j.1$ : $B$ receives $\{Y.A.B\}_{k_b}$. This message unifies only with the message $\{N_a^i.A.B\}_{k_b}$, which is the regular message that $B$ is expecting;

\item In the step $j.3$ :  $B$ receives  $\{N_b^j.B.A.Y\}_{k_b}$. This message unifies only with $\{X.B.A.N_a^i\}_{k_b}$. Upon replacing $X$ by $N_b^j$ and $Y$ by $N_a^i$, the received message becomes  $\{N_b^j.B.A.N_a^i\}_{k_b}$, which is the regular message that $B$ is expecting.
\end{enumerate}

Therefore, this protocol is a tagged one and Theorem  TSTP applies.

\subsection{Analyzing the generalized role of  $A$}

As defined in the generalized role $\cal{A}_{\cal{G}}$, an agent $A$ may participate in two  receiving-sending steps. In the first step, he receives nothing and sends the message $\{N_a^i.A.B\}_{k_b}$. In the subsequent step, he receives the message $\{A.B.N_a^i\}_{k_a}.\{B.A.X\}_{k_a}$ and sends the message $\{X.B.A.N_a^i\}_{k_b}$. This is represented by the following two rules.
\[{S_{A}^{1}}:\frac{\Box}{\{N_a^i.A.B\}_{k_b}} ~~~~~~~~~~~~~~~~~~~~~~~~ {S_{A}^{2}}:\frac{\{A.B.N_a^i\}_{k_a}.\{B.A.X\}_{k_a}}{\{X.B.A.N_a^i\}_{k_b}}\]

\subsubsection{Analyzing exchanged messages in $S_{A}^{1}$}
$ $\\
$ $\\
1- For $N_a^i$:$ $\\
$ $\\
a- On sending: $r_{S_{A}^{1}}^+=\{N_a^i.A.B\}_{k_b}$ \\
 
\scalebox{0.99}{
\begin{tabular}{lcll}
$F'(N_a^i,r_{S_{A}^{1}}^+)$ & $=$ & $F'(N_a^i,\{N_a^i.A.B\}_{k_b})$& \\
&  &\!\!\!\!\!\!\!\!\!\!\!\!\!\!\!\!\!\!\!\!\!\!\!\!\!\!\!\!\!\!\!\!\!\!\!\!\!\!\!\!\!\!\!\!\!\!\!\!\!\!\!\!\!\!\!\!\!\!\!\!\!\!\!\!\!\!\!\!\!\!\{No variable in the neighborhood of $N_a^i$ to be removed by derivation\}  &  \\
 & = & $F(N_a^i,\{N_a^i.A.B\}_{k_b})$ & \\
 &   & \{Definition \ref{reliable}\}&  \\
 & = & $\{A, B\} \cup \ulcorner k_b^{-1}\urcorner$ & \\
 &   & \{Since $\ulcorner k_b^{-1}\urcorner$ = \{B\}\}&  \\
 & = & $\{A, B\} \cup \{B\}$ & \\
 & = & $\{A, B\}$~~~~~~~~~~~~~~~~~~~~ (1.1)& \\
\end{tabular}
}
$ $\\
b- On receiving: $R_{S^{i}}^-=\emptyset$\\
$ $\\
\scalebox{0.99}{
\begin{tabular}{lcll}
$F'(N_a^i,R_{S_{A}^{1}}^-)$ & $=$ & $F'(N_a^i,\emptyset)$& \\
&  &\!\!\!\!\!\!\!\!\!\!\!\!\!\!\!\!\!\!\!\!\!\!\!\!\!\!\!\!\!\!\!\!\!\!\!\!\!\!\!\!\!\!\!\!\!\!\!\!\!\!\!\!\!\!\!\!\!\!\!\!\!\!\!\!\!\{No variable in the neighborhood of $N_a^i$ to be removed by derivation\}  &  \\
 & = & $F(N_a^i,\emptyset)$ & \\
 &   & \{Definition \ref{reliable}\}&  \\
 & = & $\top$~~~~~~~~~~~~~~~~~~~~~~~~~~~~~~(1.2)& \\
\end{tabular}
}
$ $\\
2- Accordance with Theorem TSTP:\\
$ $\\
From (1.2) and since $\ulcorner N_a\urcorner= \{A,B\}$, we have: \\
$ $\\
$\ulcorner N_a^i\urcorner \sqcap F'(N_a^i,R_{S_{A}^{1}}^-)  =\{A,B\} \sqcap \top =\{A,B\}$ ~~(1.3)\\
$ $\\
From (1.1) and (1.3), we have : \\
$ $\\
$F'(N_a^i,r_{S_{A}^{1}}^+) \sqsupseteq \ulcorner N_a^i\urcorner \sqcap F'(N_a^i,R_{S_{A}^{1}}^-)$ ~~~~~~~~~~~~~~(1.4)\\
$ $\\
From  (1.4), $S_{A}^{1}$ respects Theorem TSTP.  ~~~~~~~~~~~~~(I)\\

\subsubsection{Analyzing exchanged messages in $S_{A}^{2}$}
$ $\\$ $\\
1- For $N_a^i$:$ $\\
$ $\\
a- On sending: $r_{S_{A}^{2}}^+=\{X.B.A.N_a^i\}_{k_b}$ \\
$ $\\
\scalebox{0.99}{
\begin{tabular}{lcll}
$F'(N_a^i,r_{S_{A}^{2}}^+)$ & $=$ & $F'(N_a^i,\{X.B.A.N_a^i\}_{k_b})$& \\
&  & \{The variable $X$ is removed by derivation\}  &  \\
 & = & $F(N_a^i,\{B.A.N_a^i\}_{k_b})$ & \\
 &   & \{Definition \ref{reliable}\}&  \\
 & = & $\{A, B\} \cup \ulcorner k_b^{-1}\urcorner$ & \\
 &   & \{Since $\ulcorner k_b^{-1}\urcorner$ = \{B\}\}&  \\
 & = & $\{A, B\} \cup \{B\}$ & \\
 & = & $\{A, B\}$~~~~~~~~~~~~~~~~~~~~~~~~~~~~~~(2.1)& \\
\end{tabular}
}
$ $\\
b- On receiving: $R_{S_{A}^{2}}^-=\{A.B.N_a^i\}_{k_a}.\{B.A.X\}_{k_a}$\\
$ $\\
\scalebox{0.99}{
\begin{tabular}{lcll}
$F'(N_a^i,R_{S_{A}^{2}}^-)$ & $=$ & $F'(N_a^i,\{A.B.N_a^i\}_{k_a}.\{B.A.X\}_{k_a})$& \\
&  & \{The variable $X$ is removed by derivation\}  &  \\
 & = & $F(N_a^i,\{A.B.N_a^i\}_{k_a}.\{B.A\}_{k_a})$ & \\
 &   & \{Definition \ref{reliable}  and $F$ is well-formed\}&  \\
 & = & $F(N_a^i,\{A.B.N_a^i\}_{k_a}) \sqcap F(N_a^i,\{B.A\}_{k_a})$ & \\
 &   & \{$F$ is well-formed\}&  \\
 & = & $F(N_a^i,\{A.B.N_a^i\}_{k_a}) \sqcap \top$ & \\
 &   & \{Security lattice property\}&  \\
 & = & $F(N_a^i,\{A.B.N_a^i\}_{k_a})$ & \\
 &   & \{Definition \ref{reliable}\}&  \\
 & = & $\{A, B\} \cup \ulcorner k_a^{-1}\urcorner$ & \\
 &   & \{Since $\ulcorner k_a^{-1}\urcorner$ = \{A\}\}&  \\
 & = & $\{A, B\} \cup \{A\}$ & \\
 & = & $\{A, B\}$~~~~~~~~~~~~~~~~~~~~~~~~~~~~~(2.2)& \\
\end{tabular}
}
$ $\\
2- For $X$:$ $\\
$ $\\
c- On sending: $r_{S_{A}^{2}}^+=\{X.B.A.N_a^i\}_{k_b}$ \\
$ $\\
\scalebox{0.99}{
\begin{tabular}{lcll}
$F'(X,r_{S_{A}^{2}}^+)$ & $=$ & $F'(X,\{X.B.A.N_a^i\}_{k_b})$& \\
&  &\!\!\!\!\!\!\!\!\!\!\!\!\!\!\!\!\!\!\!\!\!\!\!\!\!\!\!\!\!\!\!\!\!\!\!\!\!\!\!\!\!\!\!\!\!\!\!\!\!\!\{No variable in the neighborhood of $X$ to be removed by derivation\}  &  \\
 & = & $F(X,\{X.B.A.N_a^i\}_{k_b})$ & \\
 &   & \{Definition \ref{reliable}\}&  \\
 & = & $\{A, B\} \cup \ulcorner k_b^{-1}\urcorner$ & \\
 &   & \{Since $\ulcorner k_b^{-1}\urcorner$ = \{B\}\}&  \\
 & = & $\{A, B\} \cup \{B\}$ & \\
 & = & $\{A, B\}$~~~~~~~~~~~~~~~~~~~~~~~~~~~~~~~(2.3)& \\
\end{tabular}
}
$ $\\
d- On receiving: $R_{S_{A}^{2}}^-=\{A.B.N_a^i\}_{k_a}.\{B.A.X\}_{k_a}$\\
$ $\\
\scalebox{0.99}{
\begin{tabular}{lcll}
$F'(X,R_{S_{A}^{2}}^-)$ & $=$ & $F'(X,\{A.B.N_a^i\}_{k_a}.\{B.A.X\}_{k_a})$& \\
&  &\!\!\!\!\!\!\!\!\!\!\!\!\!\!\!\!\!\!\!\!\!\!\!\!\!\!\!\!\!\!\!\!\!\!\!\!\!\!\!\!\!\!\!\!\!\!\!\!\!\!\!\!\!\!\{No variable in the neighborhood of $X$ to be removed by derivation\}  &  \\
 & = & $F(X,\{A.B.N_a^i\}_{k_a}.\{B.A.X\}_{k_a})$ & \\
 &   & \{Definition \ref{reliable}  and $F$ is well-formed\}&  \\
 & = & $F(X,\{A.B.N_a^i\}_{k_a}) \sqcap F(X,\{B.A.X\}_{k_a})$ & \\
 &   & \{$F$ is well-formed\}&  \\
 & = & $\top \sqcap F(X,\{B.A.X\}_{k_a})$ & \\
 &   & \{Security lattice property\}&  \\
 & = & $F(X,\{B.A.X\}_{k_a})$ & \\
 &   & \{Definition \ref{reliable}\}&  \\
 & = & $\{A, B\} \cup \ulcorner k_a^{-1}\urcorner$ & \\
 &   & \{Since $\ulcorner k_a^{-1}\urcorner$ = \{A\}\}&  \\
 & = & $\{A, B\} \cup \{A\}$ & \\
 & = & $\{A, B\}$~~~~~~~~~~~~~~~~~~~~~~~~~~~~~~(2.4)& \\
\end{tabular}
}
$ $\\
3- Accordance with Theorem  TSTP:\\
$ $\\
From (2.1),  (2.2), we have directly:\\
$$F'(N_a^i,r_{S_{A}^{2}}^+) \sqsupseteq \ulcorner N_a^i\urcorner \sqcap F'(N_a^i,R_{S_{A}^{2}}^-) ~~~~~~~~~~\mbox{(2.5)}$$
From (2.3) and  (2.4), we have directly:\\
$$F'(X,r_{S_{A}^{2}}^+) \sqsupseteq \ulcorner X\urcorner \sqcap F'(X,R_{S_{A}^{2}}^-) ~~~~~~~~~~~~~~\mbox{(2.6)}$$
From  (2.5) and (2.6), $S_{A}^{2}$ respects Theorem TSTP.  ~~~~~~(II)

\subsection{Analyzing the generalized role of  $B$}

As defined in the generalized role $\cal{B}_{\cal{G}}$, an agent $B$ may participate in just one  receiving-sending step in which he receives the message $\{Y.A.B\}_{k_b}$ and sends the message $\{A.B.Y\}_{k_a}. \{B.A.N_b^j\}_{k_a}$. This is represented by the following rule.
\[{S_{B}}:\frac{\{Y.A.B\}_{k_b}}{\{A.B.Y\}_{k_a}. \{B.A.N_b^j\}_{k_a}}\]
\subsubsection{Analyzing exchanged messages in $S_{B}$}
$ $\\$ $\\
1- For $N_b^j$:$ $\\
$ $\\
a- On sending: $r_{S_{B}}^+=\{A.B.Y\}_{k_a}. \{B.A.N_b^j\}_{k_a}$ \\
$ $\\
\scalebox{0.99}{
\begin{tabular}{lcll}
$F'(N_b^j,r_{S_{B}}^+)$ & $=$ & $F'(N_b^j,\{A.B.Y\}_{k_a}. \{B.A.N_b^j\}_{k_a})$& \\
&  & \{The variable $Y$ is removed by derivation\}  &  \\
 & = & $F(N_b^j,\{A.B\}_{k_a}. \{B.A.N_b^j\}_{k_a})$ & \\
 &   & \{Definition \ref{reliable}  and $F$ is well-formed\}&  \\
 & = & $F(N_b^j,\{A.B\}_{k_a}) \sqcap F(N_b^j, \{B.A.N_b^j\}_{k_a})$ & \\
&   & \{Definition \ref{reliable}  and $F$ is well-formed\}&  \\
 & = & $F(N_b^j,\{A.B\}_{k_a}) \sqcap F(N_b^j,\{B.A.N_b^j\}_{k_a})$ & \\
 &   & \{$F$ is well-formed\}&  \\
 & = & $ \top \sqcap F(N_b^j,\{B.A.N_b^j\}_{k_a}) $ & \\
 &   & \{Security lattice property\}&  \\
 & = & $F(N_b^j,\{B.A.N_b^j\}_{k_a})$ & \\

 &   & \{Definition \ref{reliable}\}&  \\
 & = & $\{A, B\} \cup \ulcorner k_a^{-1}\urcorner$ & \\
 &   & \{Since $\ulcorner k_a^{-1}\urcorner$ = \{A\}\}&  \\
 & = & $\{A, B\} \cup \{A\}$ & \\
 & = & $\{A, B\}$~~~~~~~~~~~~~~~~~~~~~~~~~(3.1)& 
\end{tabular}
}
$ $\\
b- On receiving: $R_{S_{B}}^-=\{Y.A.B\}_{k_b}$\\
$ $\\
\scalebox{0.99}{
\begin{tabular}{lcll}
$F'(N_b^j,R_{S_{B}}^-)$ & $=$ & $F'(N_b^j,\{Y.A.B\}_{k_b})$& \\
&  & \{The variable $Y$ is removed by derivation\}  &  \\
 & = & $F(N_b^j,\{A.B\}_{k_b})$ & \\
 &   & \{$F$ is well-formed\}&  \\
 & = & $\top$~~~~~~~~~~~~~~~~~~~~~~~~~~~~~~~(3.2)& \\
\end{tabular}
}
$ $\\
2- For $Y$:$ $\\
$ $\\
a- On sending: $r_{S_{B}}^+=\{A.B.Y\}_{k_a}. \{B.A.N_b^j\}_{k_a}$ \\
$ $\\
\scalebox{0.99}{
\begin{tabular}{lcll}
$F'(Y,r_{S_{B}}^+)$ & $=$ & $F'(Y,\{A.B.Y\}_{k_a}. \{B.A.N_b^j\}_{k_a})$& \\
&  &\!\!\!\!\!\!\!\!\!\!\!\!\!\!\!\!\!\!\!\!\!\!\!\!\!\!\!\!\!\!\!\!\!\!\!\!\!\!\!\!\!\!\!\!\!\{No variable in the neighborhood of $Y$ to be removed by derivation\}  &  \\
 & = & $F(Y,\{A.B.Y\}_{k_a}. \{B.A.N_b^j\}_{k_a})$ & \\
 &   & \{Definition \ref{reliable}  and $F$ is well-formed\}&  \\
 & = & $F(Y,\{A.B.Y\}_{k_a}) \sqcap F(Y,\{B.A.N_b^j\}_{k_a})$ & \\
 &   & \{$F$ is well-formed\}&  \\
 & = & $   F(Y,\{A.B.Y\}_{k_a}) \sqcap \top$ & \\
 &   & \{Security lattice property\}&  \\
 & = & $   F(Y,\{A.B.Y\}_{k_a})$ & \\
 &   & \{Definition \ref{reliable}\}&  \\
 & = & $\{A, B\} \cup \ulcorner k_a^{-1}\urcorner$ & \\
 &   & \{Since $\ulcorner k_a^{-1}\urcorner$ = \{A\}\}&  \\
 & = & $\{A, B\} \cup \{A\}$ & \\
 & = & $\{A, B\}$~~~~~~~~~~~~~~~~~~~~~~~~~~~(3.3)& \\
\end{tabular}
}
$ $\\
b- On receiving: $R_{S_{B}}^-=\{Y.A.B\}_{k_b}$\\
$ $\\
\scalebox{0.99}{
\begin{tabular}{lcll}
$F'(Y,R_{S_{B}}^-)$ & $=$ & $F'(Y,\{Y.A.B\}_{k_b})$& \\
&  &\!\!\!\!\!\!\!\!\!\!\!\!\!\!\!\!\!\!\!\!\!\!\!\!\!\!\!\!\!\!\!\!\!\!\!\!\!\!\!\!\!\!\!\!\!\!\!\!\!\!\!\!\!\{No variable in the neighborhood of $Y$ to be removed by derivation\}  &  \\
 & = & $F(Y,\{Y.A.B\}_{k_b})$ & \\
&   & \{Definition \ref{reliable}\}&  \\
 & = & $\{A, B\} \cup \ulcorner k_b^{-1}\urcorner$ & \\
 &   & \{Since $\ulcorner k_b^{-1}\urcorner$ = \{B\}\}&  \\
 & = & $\{A, B\} \cup \{B\}$ & \\
 & = & $\{A, B\}$~~~~~~~~~~~~~~~~~~~~~~~~~~(3.4)& \\
\end{tabular}
}
$ $\\
3- Accordance with Theorem  TSTP:\\
$ $\\
From (3.1),  (3.2) and since $\ulcorner N_b\urcorner=\{A, B\}$  we have:\\
$$F'(N_b^j,r_{S_{B}}^+) \sqsupseteq \ulcorner N_b^j\urcorner \sqcap F'(N_b^j,R_{S_{B}}^-) ~~~~~~~~~~~~~~~~\mbox{(3.5)}$$
From (3.3) and  (3.4), we have directly:\\
$$F'(Y,r_{S_{B}}^+) \sqsupseteq \ulcorner Y\urcorner \sqcap F'(Y,R_{S_{A}}^-) ~~~~~~~~~~~~~~~~~~~~~\mbox{(3.6)}$$
From  (3.5) and (3.6), $S_{B}$ respects Theorem TSTP.  ~~~~~~~~(III)

\normalsize
\section{Comparison with related works}\label{sec5}

From (I), (II) and (II), we deduce that the tagged version of the Needham-Schroeder  public-key protocol given in Table  \ref{NSLVar} fully respects Theorem TSTP. Hence, we conclude that it is correct for secrecy. In fact, tagging constitutes an efficient way to create well-structured protocols that help avoid misinterpretation of received messages and a regular agent is always assured that he is receiving messages from the right regular agent. A tagged protocol is a good candidate for an analysis by witness-functions that can verify it quickly owing to the simplified theorem we have exhibited so far. By the same token, the authors in \cite{TAG1} add a tag for each type by adding an explicit name in every message generated by the protocol. For instance, they use the notation (nonce, $N$) to indicate that the value $N$ is supposed to be a nonce. This extra information is in fact added by honest agents to precise the intended type of the message and the receiver uses it to recognize the message. This way, tags ensure that any message having originally a given type will not be interpreted as having another type which prevents any possible type-flaw attack. In \cite{TAG2}, tagging schemes are used for a decidability proof purpose. The tag is represented as a fresh number that marks all encrypted sub-terms in the protocol. As a result, tagging prevents the unification of different encrypted sub-terms which transforms an undecidable general problem to a decidable particular one even with an infinite number of nonces. In \cite{TAG3}, tagging allows to change the inherent non-termination property caused by inference rules. An approach based on Horn clauses \cite{blanchet:hal-01110425} is adopted in which attacker abilities and protocol rules are translated into Horn clauses, then the algorithm infers progressively new clauses by resolution. After some resolution steps, the authors show that it is possible to generate an infinite number of sessions which may lead to non-termination. However, after adding a tag on each use of a cryptographic primitive, every encrypted message becomes distinguishable from others, obviously owing to the tag. To practically highlight the effect of tagging, they apply their approach on untagged protocols whose their resolution algorithm does not terminate (i.e. the Needham–Schroeder shared-key protocol, the Woo-Lam shared key protocol, etc.). Then, they show that after tagging the protocol, messages become unambiguously identified and the infinite loop observed before never happens again. Therefore, the algorithm terminates. In \cite{Arapinis}, Arapinis et al. give a scheme to transform a secure protocol for a single session, which is a decidable problem, to a secure one for an unbounded number of sessions using tagging. In \cite{10.1007/978-3-540-77050-3_29}, Cortier et al. show that if a protocol running alone is secure, it remains secure even if it runs  simultaneously with other protocols if we carefully add a tag to every encryption in such a way that we can differentiate between all the protocols by adding the name of the protocol for example. Similarly,  Bauer et al. \cite{10.1007/978-3-662-49635-0_10} show that if a protocol is correct for secrecy with a probability higher than some threshold, a protocol composition remains correct for secrecy provided that protocol messages are tagged. Our work in this paper is one of these efforts with the clear advantage that our proposed theorem helps to prove secrecy statically with no need to go through dynamic complexities.

\section{Conclusion}\label{sec6}

In this paper, we  put forward a new theorem to prove secrecy inside tagged protocols using witness-functions. Then, we run a detailed analysis on a tagged version of the Needham-Schroeder  public-key protocol. Finally, we discussed some works pinpointing multiple advantages of protocol tagging.



\bibliographystyle{ieeetr}

\tiny

\bibliography{Ma_these}

\end{document}